\documentclass[twocolumn]{webofc}

\usepackage[varg]{txfonts}
\usepackage[pdftex,bookmarksopen]{hyperref} 
\usepackage{multirow}

\newcommand{\nn}{\nonumber}

\begin{document}

\title{Towards a combined analysis of inclusive/exclusive electroproduction}

\author{\firstname{Astrid N.} \lastname{Hiller Blin}\inst{1}\fnsep\thanks{\email{hillerbl@uni-mainz.de}} \and
        \firstname{Vitaly V.} \lastname{Chesnokov}\inst{3}\fnsep \and
        \firstname{Victor I.} \lastname{Mokeev}\inst{2}\fnsep
}

\institute{Institut f\"ur Kernphysik \& PRISMA$^+$ Cluster of Excellence,
Johannes Gutenberg Universit\"at, D-55099 Mainz, Germany
\and
          Thomas Jefferson National Accelerator Facility, Newport News, VA 23606, USA
\and
           Skobeltsyn Nuclear Physics Institute and Physics Department at Moscow State University, 119899 Moscow, Russia
          }

\abstract{
In view of the major advances achieved by the CLAS experiments  in studying the $N^*$  electroexcitation amplitudes, as well as further extension of these studies in the experiments with CLAS12, we present an approach for the evaluation of the resonant contributions to inclusive electron scattering off protons. For the first time, the resonant contributions are determined from the experimental results on $N^*$ electrocouplings available from the data analyses of exclusive meson electroproduction off protons. This is a useful benchmark for future endeavours on understanding the transition
region between low and high-energy regions, strongly related to tests on quark-hadron duality.
}
\maketitle
\section{Introduction}
\label{intro}

Due to their connection to the nucleon parton distribution functions (PDFs), which were recently related to the QCD Lagrangian (see Refs.~\cite{Lin:2017snn,Jimenez-Delgado:2013sma,Gao:2017yyd} and references therein for an overview), and in view of the exploration of  the quark-hadron duality ~\cite{Bloom:1970xb}, the interest in inclusive electron scattering reactions off nucleons has been strong. The 
 Jefferson Lab (JLab) data played an important role in these studies~\cite{Osipenko:2003bu,Malace:2009kw,Christy:2007ve,JLab:E00-002,Liang:2004tj}. The experiments in Halls A/C \cite{Malace:2009kw,Christy:2007ve} provided precise information on inclusive electron scattering at the invariant masses of the final hadrons of $W<4.0$~GeV and photon virtualities $Q^2<6.0$~GeV$^2$ allowing us to disentangle the transverse and longitudinal parts of the inclusive cross sections. The CLAS detector offers a unique possibility of obtaining the inclusive structure functions $F_2$ in the broad range of $W<2.5$~GeV at given 
 photon virtuality $Q^2<4.5$~GeV$^2$~\cite{Golubenko:2019gxz}. With the CLAS12 experiments, the largest coverage ever achieved in the resonance region, of $Q^2<12$~GeV$^2$ will be reached~\cite{Burkert:2018nvj}. The broad coverage in $W$ at given $Q^2$ is particularly important in the resonance region: due to the presence of resonant structures in the observable kinematics, with an intricate behavior as functions of $W$ and $Q^2$, the interpolation over $W$ and $Q^2$ to extract the structure functions becomes challenging in the resonance region. In addition, the analyses of CLAS experiments on exclusive meson electroproduction enabled to determine the nucleon resonance $N^*$ electro-excitation amplitudes in terms of the transverse, $A_{1/2}(Q^2)$, $A_{3/2}(Q^2)$ and longitudinal, $S_{1/2}(Q^2)$  electrocouplings~\cite{Aznauryan:2011qj,Mokeev:2018zxt}. This information has become available for most excited nucleon states in the mass range $M_{N^*}<1.8$~GeV and at $Q^2<5.0$~GeV$^2$, and substantial evidence of a new baryon state $N^\prime(1720)~3/2^+$ has been found recently~\cite{Mokeev:2015moa,Burkert:2019opk}. In the near future, the data on $\pi^+\pi^-p$ electroproduction with CLAS will provide electrocouplings of most $N^*$ in the mass range up to 2.0~GeV and at $Q^2<5.0$~GeV$^2$ \cite{Mokeev:2019new}. 
 
The availability of  electrocoupling 
data on individual nucleon resonances 
makes it possible to, in this work, evaluate the 
resonant contributions to the inclusive electron scattering observables, in a fashion independent of the inclusive data, with the use of a relativistic Breit-Wigner ansatz~\cite{HillerBlin:2019hhz}. 
In a next step, together with a non-resonant background, this will allow to determine the transition from low to high energies and photon virtualities $Q^2$, and consequently between resonance bound valence quarks and asymptotically free partons.

\section{Formalism}\label{S:form}
Following the approach in Ref.~\cite{HillerBlin:2019hhz} and definitions therein, we describe the  contributions of the $N^*$ and $\Delta$ resonances to the transverse (longitudinal) inclusive virtual photon-proton cross section within the relativistic Breit-Wigner ansatz
\begin{align}
\sigma_{T(L)}^R&(W,Q^2)=\frac{\pi}{q_\gamma^2}\nn\\
&\times\sum_{N^*}(2J_r+1)\frac{M_r^2\Gamma_\text{tot}(W){\Gamma_\gamma}^{T(L)}(M_r,Q^2)}{(M_r^2-W^2)^2+M_r^2\Gamma_\text{tot}^2(W)},\label{Eq:BW}
\end{align}
where the resonance masses, spins and total hadronic decay widths are given by $M_r$, $J_r$ and $\Gamma_\text{tot}(W)$, respectively. The resonance electromagnetic decay widths to the final states, $\Gamma_\gamma^{T(L)}$ are related to the resonance electrocouplings $A_{1/2}(Q^2)$, $A_{3/2}(Q^2$, and $S_{1/2}(Q^2$  as
\begin{align}
\Gamma_\gamma^T(W=M_r,Q^2)&=\frac{q^2_{\gamma,r}(Q^2)}{\pi}\frac{2M_N}{(2J_r+1)M_r}\nn\\
&\times\left(|A_{1/2}(Q^2)|^2+|A_{3/2}(Q^2)|^2\right),\nn\\
\Gamma_\gamma^L(W=M_r,Q^2)&=2\frac{q^2_{\gamma,r}(Q^2)}{\pi}\frac{2M_N}{(2J_r+1)M_r}|S_{1/2}(Q^2)|^2,\label{Eq:EMWidths}
\end{align}
with $q_{\gamma,r}=\left.q_{\gamma} \right|_{W=M_r}$.
We use the information on the resonance electrocouplings, including the new $N^\prime(1720)~3/2^+$ state~\cite{Mokeev:2015moa,Burkert:2019opk}, from interpolation and extrapolation of the CLAS data~\cite{Aznauryan:2011qj,Mokeev:2018zxt} over $Q^2$ as described in Ref.~\cite{HillerBlin:2019hhz}. The interpolation/extrapolation tools developed by CLAS~\cite{CLAS:coups} are used for the central values of the electrocouplings from to $Q^2=0.5$~GeV$^2$ to $Q^2=5.0$~GeV$^2$, and their uncertainties are estimated from the experimental data error bars. These uncertainties are propagated to the inclusive electron scattering observables via a  bootstrap based approach. At the resonant point $W$=$M_r$ the resonance total decay widths $\Gamma_\text{tot}(W)$  were taken from the PDG~\cite{Tanabashi:2018oca}. Their $W$-dependencies were evaluated as described in Ref.~\cite{HillerBlin:2019hhz}, assuming that the centrifugal barrier penetration factor shapes the $W$-evolution of the total resonance decay widths. 
The resonant contributions to the unpolarized inclusive cross section is then given by~\cite{Christy:2007ve}
\begin{align} 
\label{xs}
\sigma_U^R(W,Q^2)&=\sigma_T^R(W,Q^2)+\epsilon_T\sigma_L^R(W,Q^2),\\
 \epsilon_T &=
 \left(1+ 2\,\frac{\nu^2+Q^2}{Q^2}\tan^2\frac{\theta_e}{2}\right)^{-1},\label{SigmaU1}
 \end{align}
with $\theta_e$ the electron scattering angle and $\nu$ the energy transfer to the target. The resonant pieces of the inclusive structure functions are computed from the cross sections in Eq.~\ref{xs} as~\cite{Drechsel:2002ar}
\begin{align}
\label{sf} 
F_1^R(W,Q^2)&=\frac{K W}{4\pi^2\alpha}\sigma_T^R,\nn\\
F_2^R(W,Q^2)&=\frac{K W}{4\pi^2\alpha}\frac{2x}{1+\frac{Q^2}{\nu^2}}
\left(\sigma_T^R +\sigma_L^R\right)\nn\\
=& \frac{K W}{4\pi^2\alpha}\frac{2x}{1+\frac{Q^2}{\nu^2}} \frac{1+R_{LT}}{1+\epsilon_T R_{LT}} \sigma_U,
\end{align}
where $x=\frac{Q^2}{2M_N\nu}$,  $K=\frac{W^2-M_N^2}{2M_N}$ and  $R_{LT}=\sigma_L/\sigma_T$. We refer to Ref.~\cite{HillerBlin:2019hhz} for further details.

A special procedure has been developed for the interpolation of the experimental results on inclusive structure functions in the region of 1.07~GeV$<W<$4.0~GeV and $Q^2<7.0$~GeV$^2$ \cite{Golubenko:2019gxz}. This procedure is also used for the interpolation of the resonant contributions. In Ref.~\cite{CLAS:SFDB}, the inclusive structure functions, virtual photon and electron scattering inclusive cross sections can be computed online for the beam energies defined by the user. The user friendly interface is available to provide the computed results both as numerical files and as sets of plots.

\section{Results and discussion}\label{S:res}

\begin{figure*}
\centering
\includegraphics[width=0.4\textwidth]{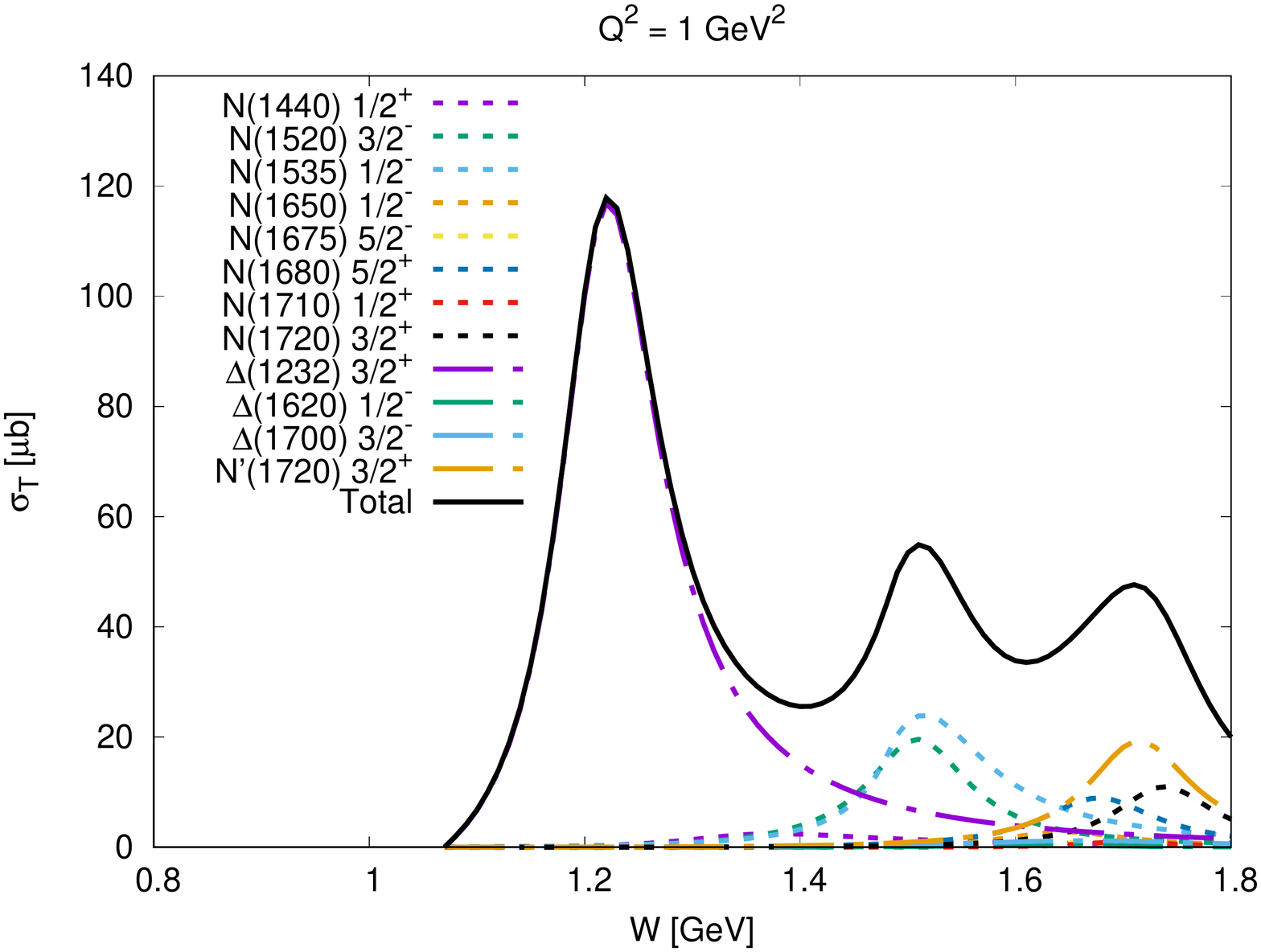}
\includegraphics[width=0.4\textwidth]{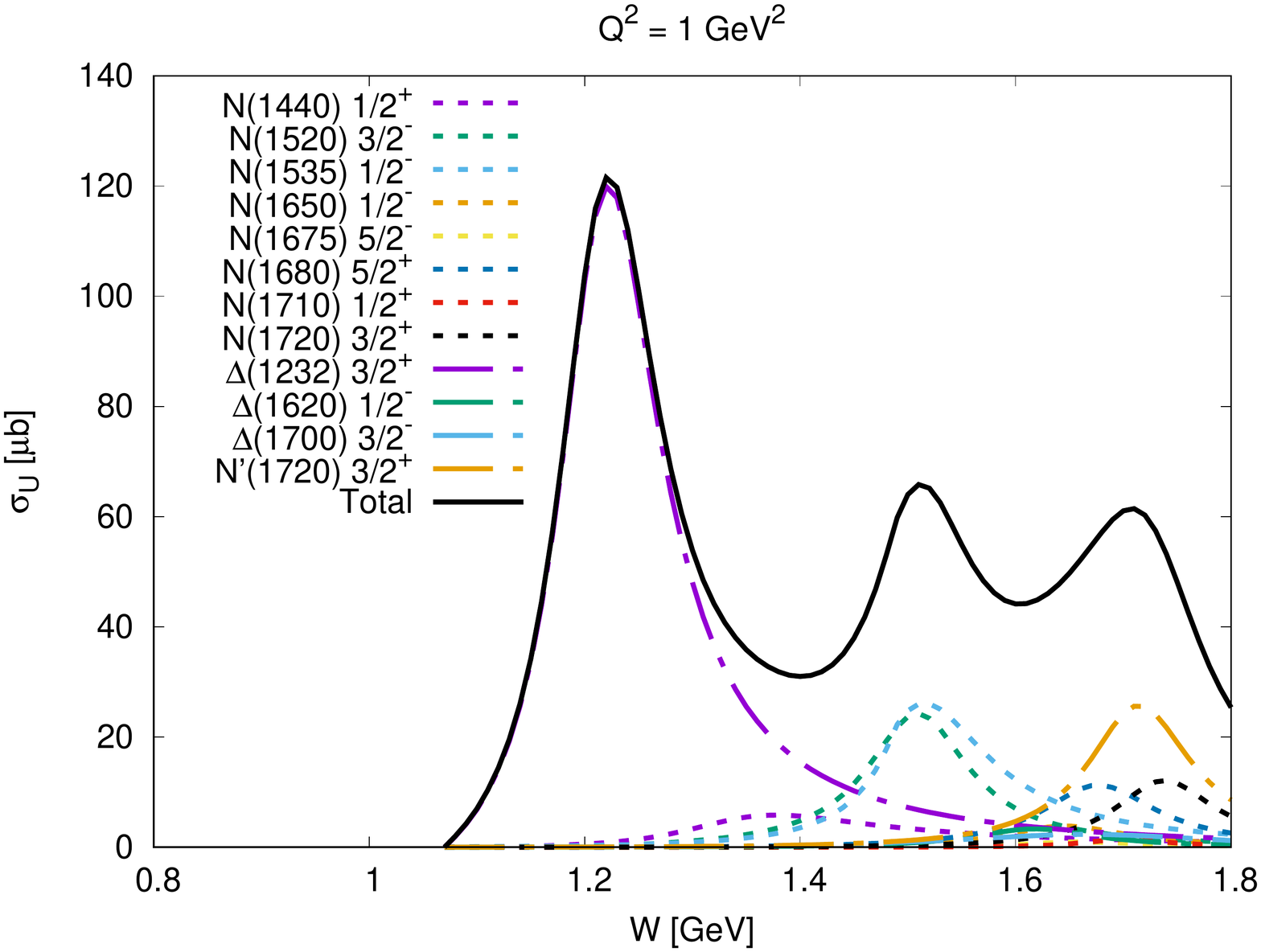}
\includegraphics[width=0.4\textwidth]{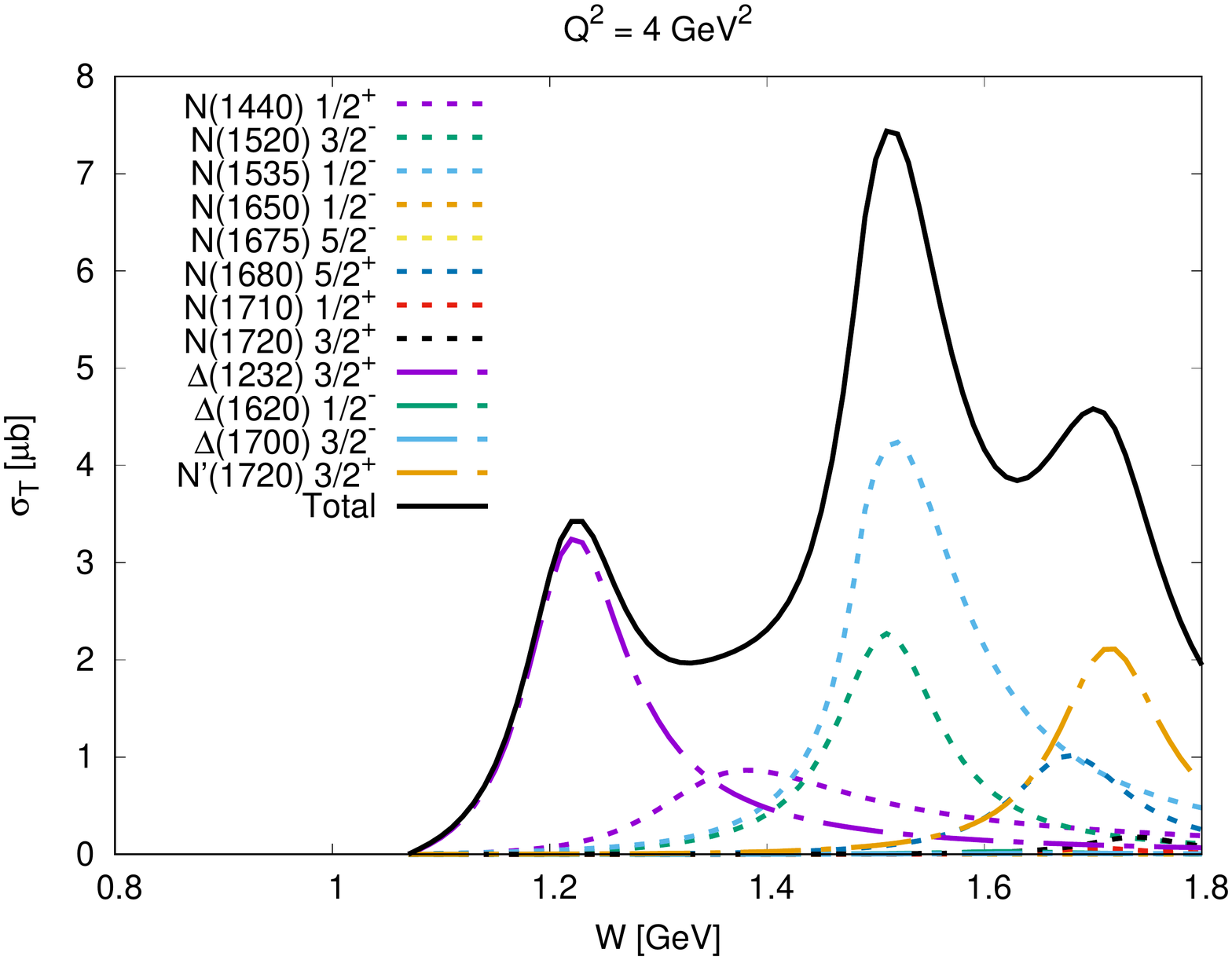}
\includegraphics[width=0.4\textwidth]{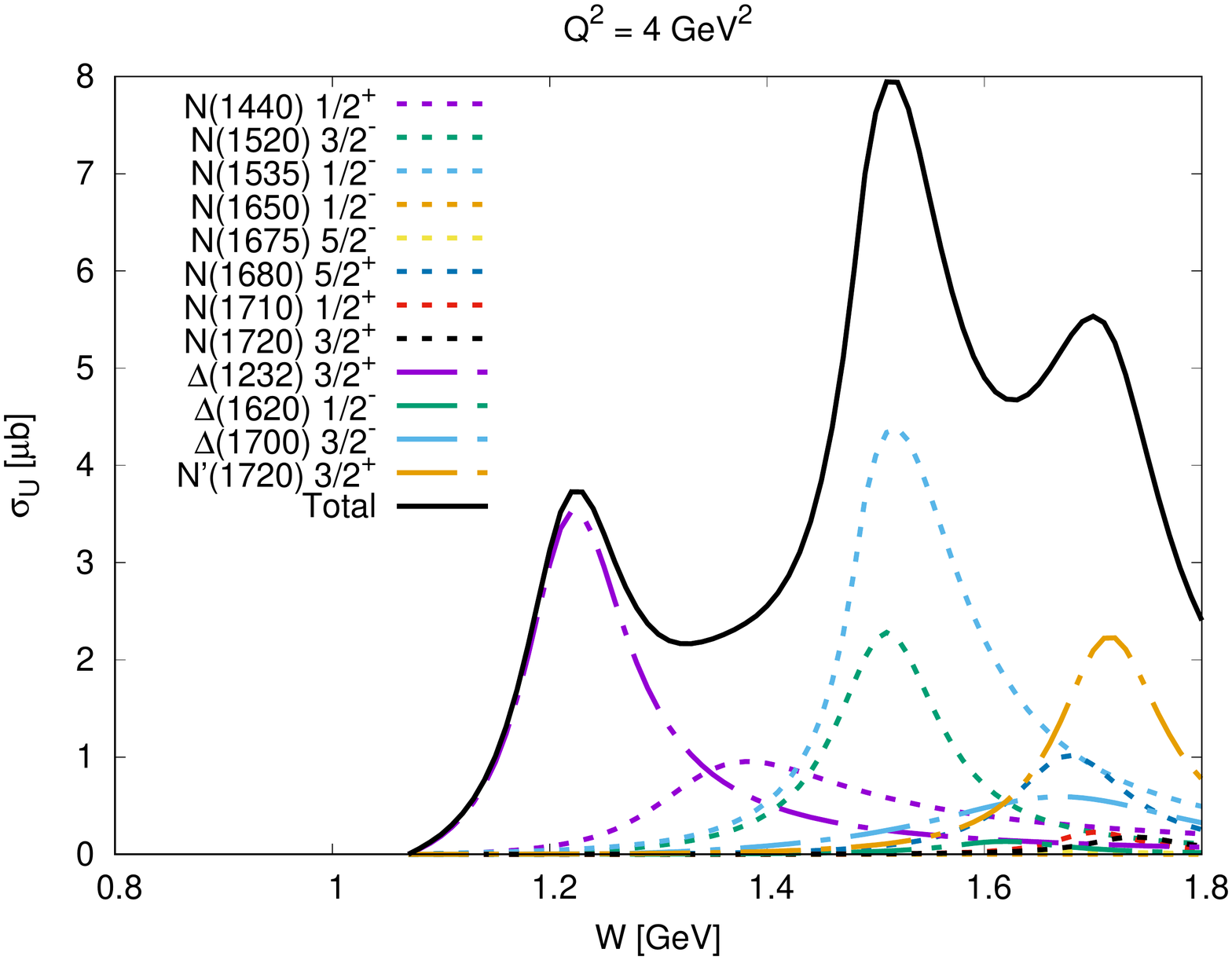}
\caption{Decomposition of the transverse (unpolarized) resonant cross sections, shown as thick black curves in the left (right) column, into the separate contributions of each resonance included in the model, at $Q^2=1$ and $4$~GeV$^2$, for an electron beam energy of 10.6~GeV.}
\label{F:SingF2TLU}
\end{figure*}
\begin{figure*}
\centering
\includegraphics[width=0.4\textwidth]{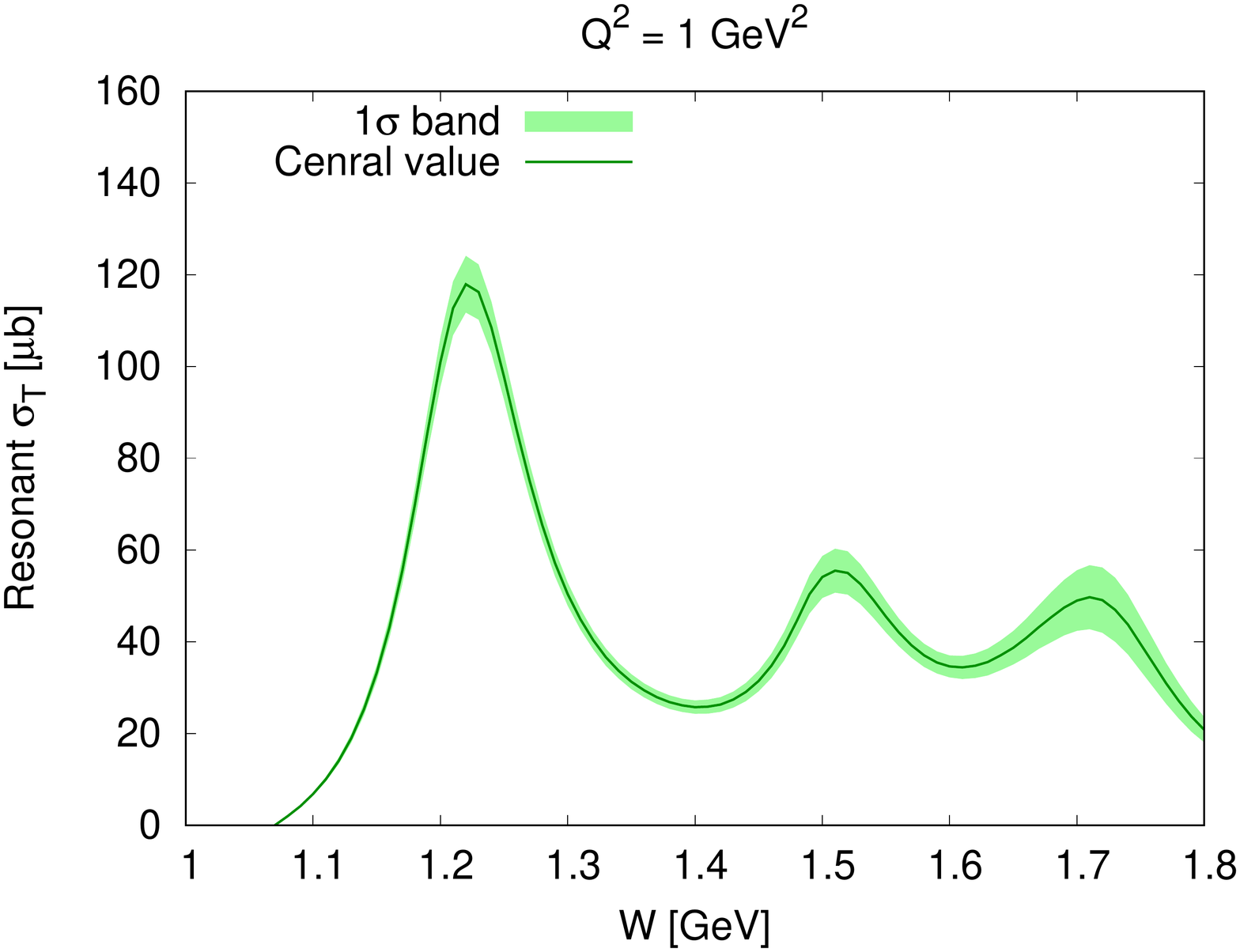}
\includegraphics[width=0.4\textwidth]{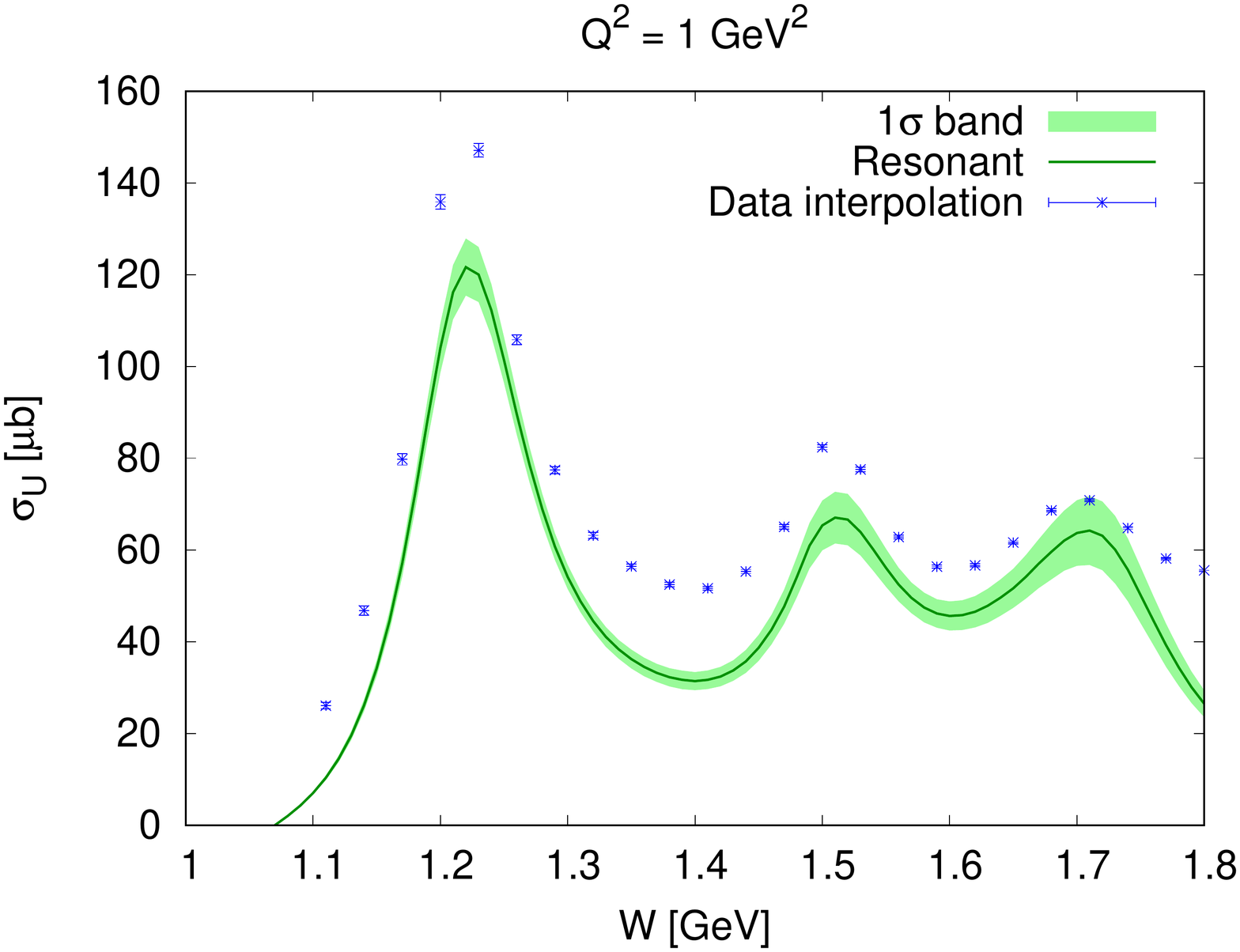}
\includegraphics[width=0.4\textwidth]{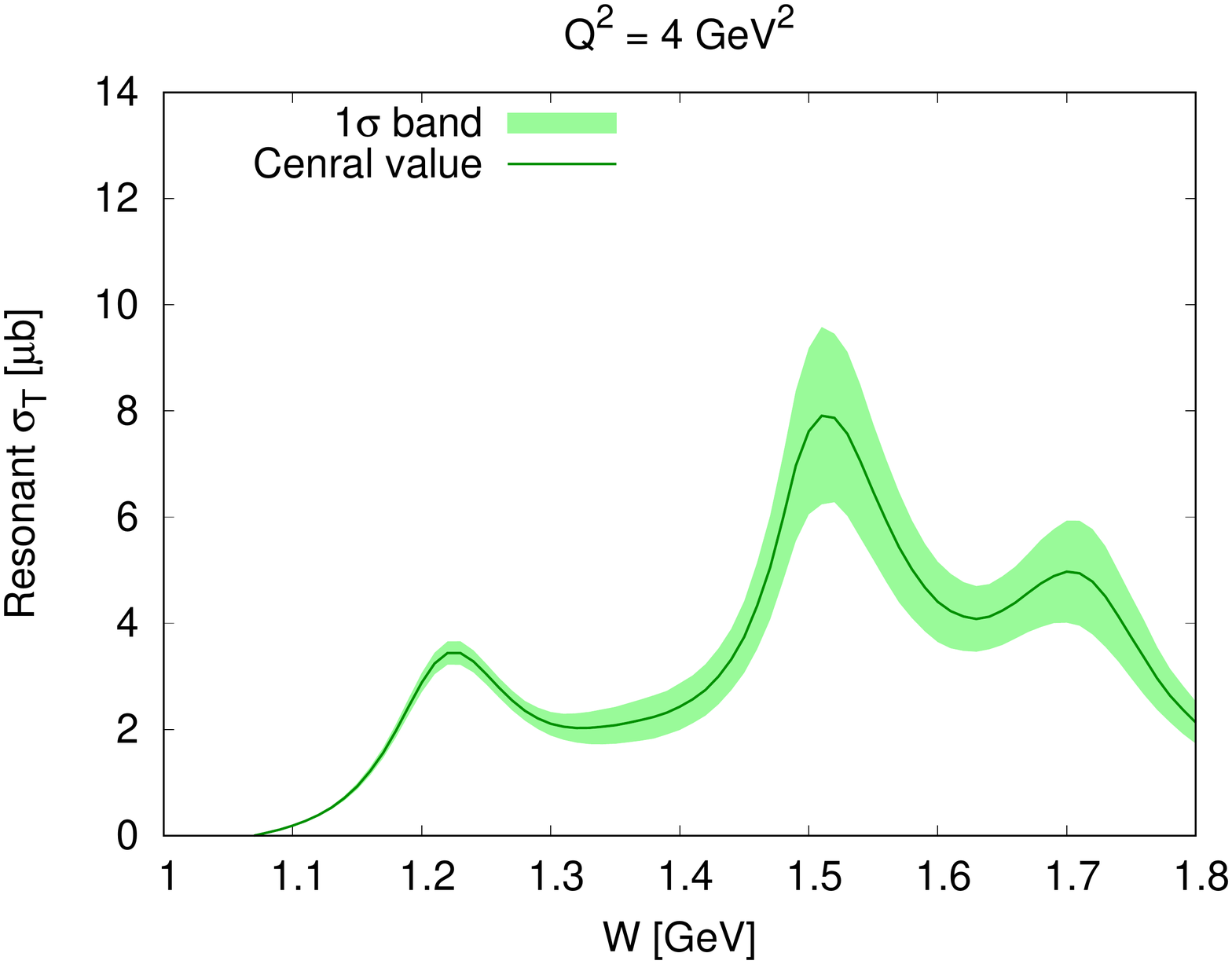}
\includegraphics[width=0.4\textwidth]{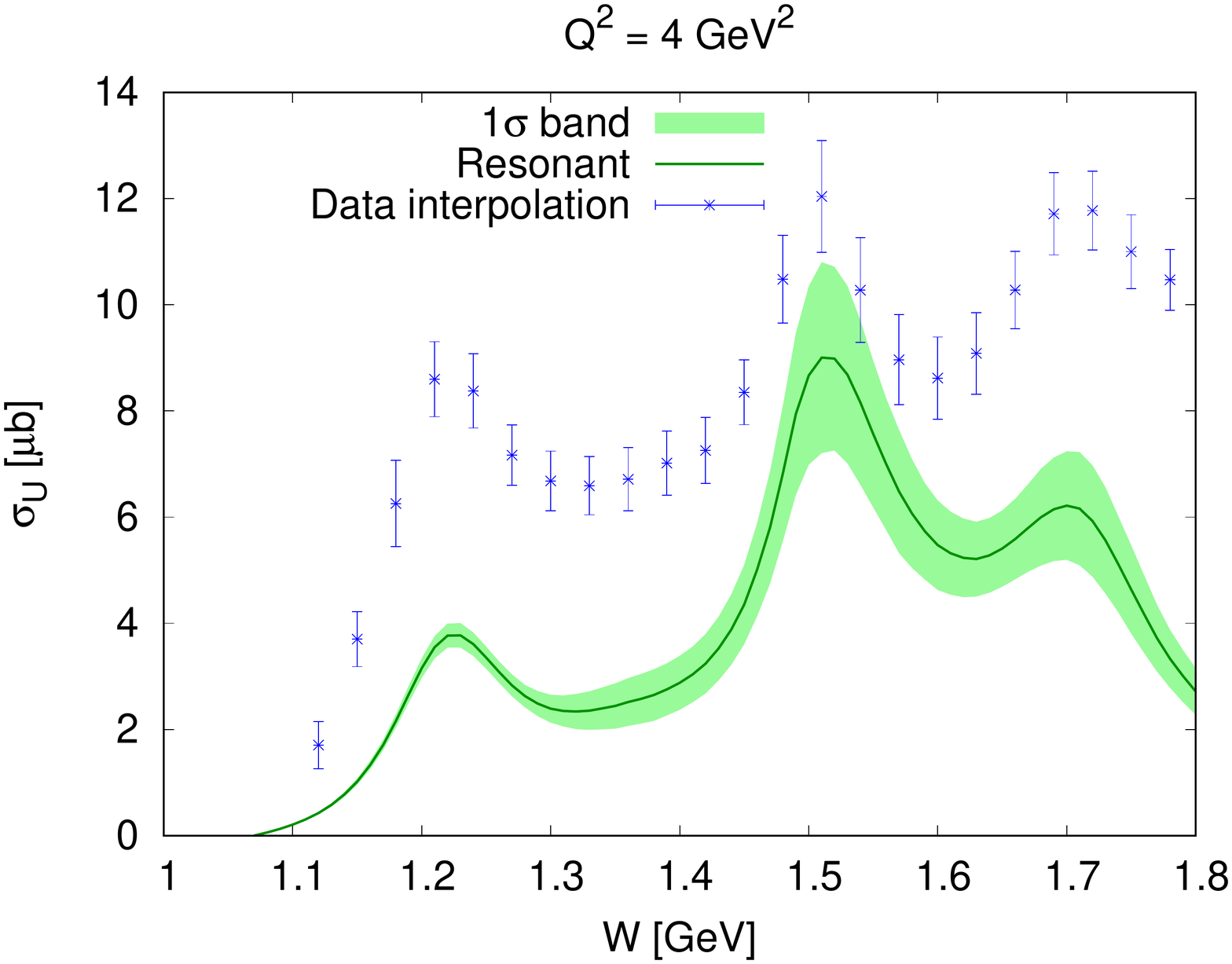}
\caption{Resonant contributions (green curves with uncertainty bands) to the transverse (unpolarized) virtual photon-proton cross sections $\sigma_{T(U)}$ are shown on the left (right) column, at $Q^2=1$ and $4$~GeV$^2$, for an electron beam energy of 10.6~GeV. The data points are from CLAS~\cite{CLAS:SFDB,Golubenko:2019gxz}.}
\label{F:sigsTLU}
\end{figure*}

In Figs.~\ref{F:SingF2TLU} and \ref{F:sigsTLU} we show the transverse (unpolarized) cross sections, $\sigma_{T(U)}^R$ for an electron beam energy of 10.6~GeV, and at representative  values of photon virtualities in the range of $Q^2\geq1$~GeV$^2$, where partons become active degrees of freedom in inclusive electron scattering. In Fig.~\ref{F:SingF2TLU}, the decomposition into the individual contributions of each resonance is shown, while in Fig.~\ref{F:sigsTLU} the resonant contribution from all resonances combined are compared with the full unpolarized cross section data~\cite{CLAS:SFDB,Golubenko:2019gxz}. Although the resonances clearly cluster into three regions, their tails give important contributions to the neighboring ones. Even the first region, clearly dominated by the single $\Delta(1232)~3/2^+$ resonance at $Q^2$ $\approx$ 1.0 GeV$^2$, is also  affected by the tail of the $N(1440)~1/2^+$. This becomes increasingly relevant at higher $Q^2$, since the transverse $\Delta(1232)~3/2^+$ electrocouplings decrease with $Q^2$ much faster than those of the $N(1440)~1/2^+$~\cite{CLAS:SFDB}. In the second resonance region, the $N(1535)~1/2^-$ gives the largest contribution at $Q^2 > 2.0$~GeV$^2$ due to the slower decrease of its  $A_{1/2}$  electrocoupling with $Q^2$ in comparison with that of $N(1520)3/2^-$~\cite{Aznauryan:2011qj,Mokeev:2018zxt}.  For the same reason, its tail gives increasingly noticeable contributions with $Q^2$ also to the third resonance region. Because of the large total decay width of the $N(1440)1/2^+$ resonance 300 MeV, its complementary contribution is observed in all three resonance regions at $W<1.7$~GeV. The contribution from several overlapping resonances combined generates the peak seen in the third resonance region of the cross sections. The largest contributions in the third region are assigned to the $N(1680)~5/2^+$ and the $N^\prime(1720)~3/2^+$ candidate. The contribution of the $N^\prime(1720)~3/2^+$ is needed in order to reproduce the third resonance maximum seen in the virtual photon inclusive cross section.

Overall, the transverse resonant part gives the largest contribution
  to the resonant cross sections. The size of the longitudinal piece increases with $W$, but does not exceed 30\% of the total cross section in the kinematical region shown here. The evolution of the unpolarized resonant cross sections with $Q^2$ is rather complex: both the first and the third  regions show a stronger fall-off with $Q^2$ than the second  peak, mainly due to the $N(1535)~1/2^-$ contribution in the second region. The different resonance states thus seem to display rather considerable differences in the $Q^2$ evolution of their electrocouplings, thus underlying the importance of dedicating separate analyses to each of them. 
  
 \begin{figure*}
\centering
\includegraphics[width=0.4\textwidth]{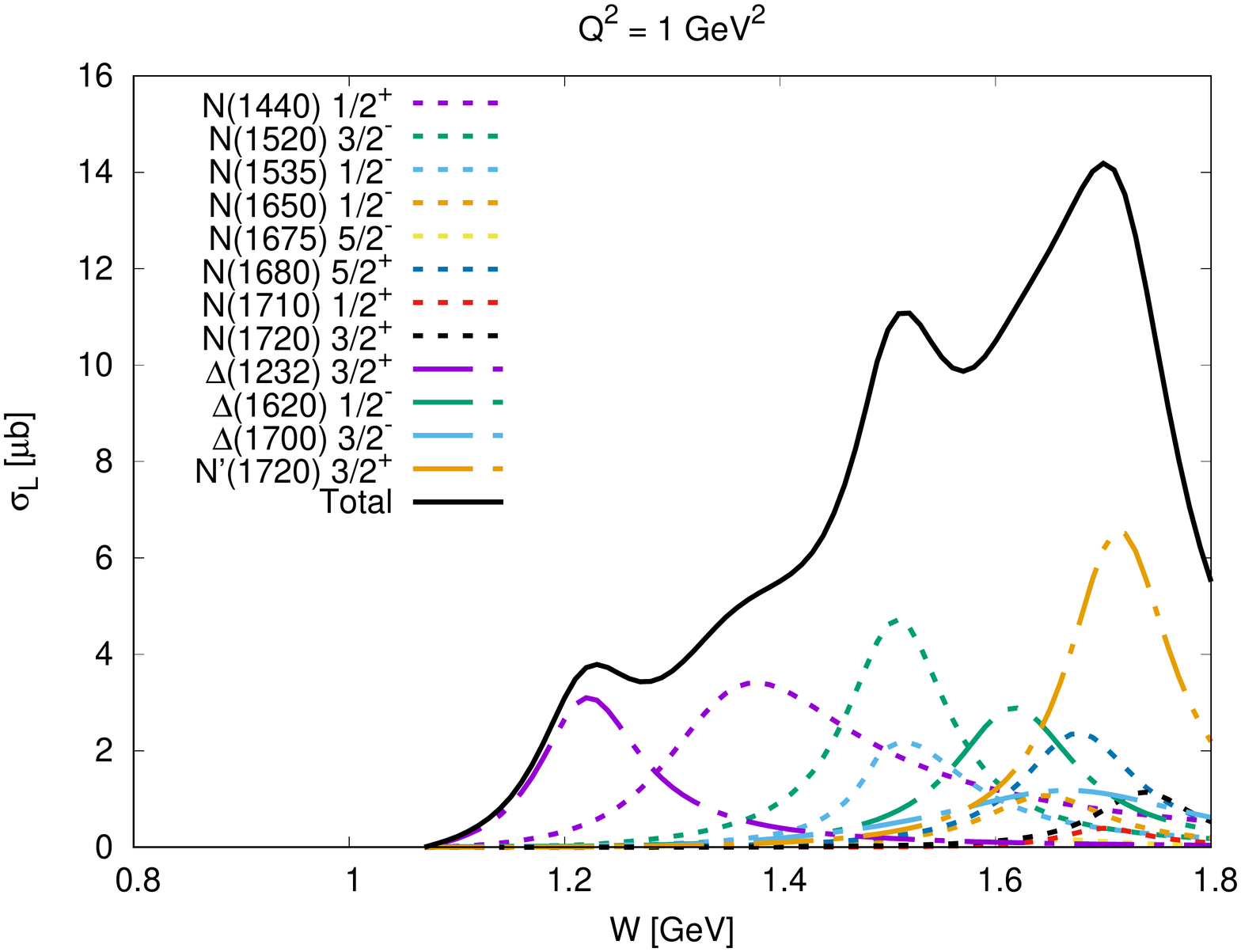}
\includegraphics[width=0.4\textwidth]{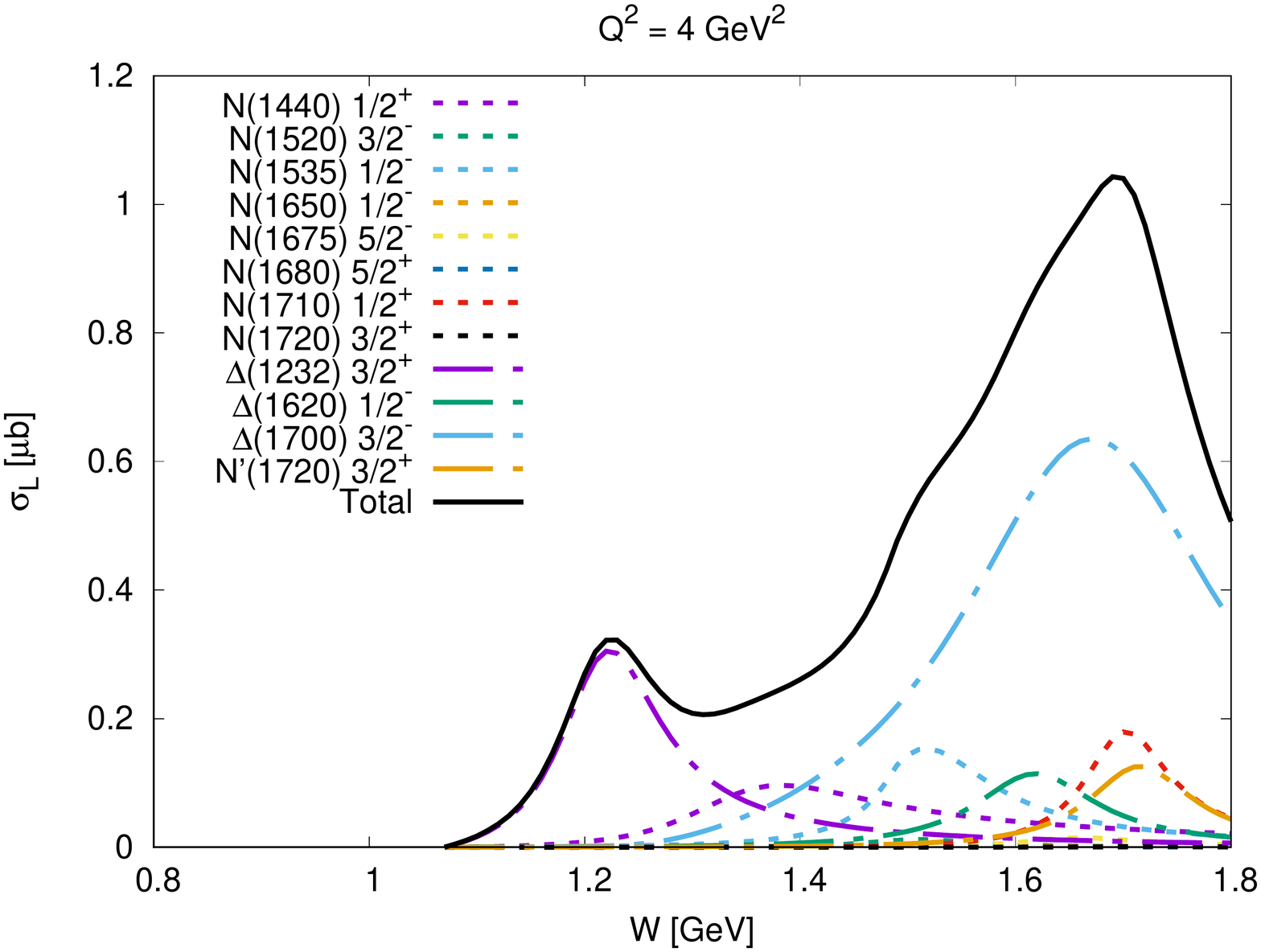}
\caption{Decomposition of the longitudinal resonant cross sections, shown as thick black curves, into the separate contributions of each resonance included in the model, at $Q^2=1$ and $4$~GeV$^2$.}
\label{F:sigsLTU}
\end{figure*} 
The predicted resonant contributions to the longitudinal virtual photon cross sections are shown in Figs.~\ref{F:sigsLTU}. The longitudinal part of the cross section is sensitive to high-lying $N^*$. The largest contributions to the longitudinal exclusive cross sections come from the resonances with masses above 1.6~GeV. The contributions from each resonance demonstrate a pronounced dependence on $Q^2$. 

Future experimental results from CLAS on resonance electrocouplings in the mass range from 1.6~GeV to 2.0~ GeV and 0.4~GeV$^2<Q^2<5.0$~GeV$^2$ expected from the data~\cite{Trivedi:2018rgo,Isupov:2017lnd,Markov:2018loh,Markov:2019fjy} will improve our estimates on the contributions from excited nucleon states in the mass range above 1.6~GeV. Furthermore, they will provide the first results on electrocouplings of the resonances in the mass range above 1.9~GeV, where the transition from the resonance to deep inelastic scattering regions takes place.

\begin{figure*}
\centering
\includegraphics[width=0.45\textwidth]{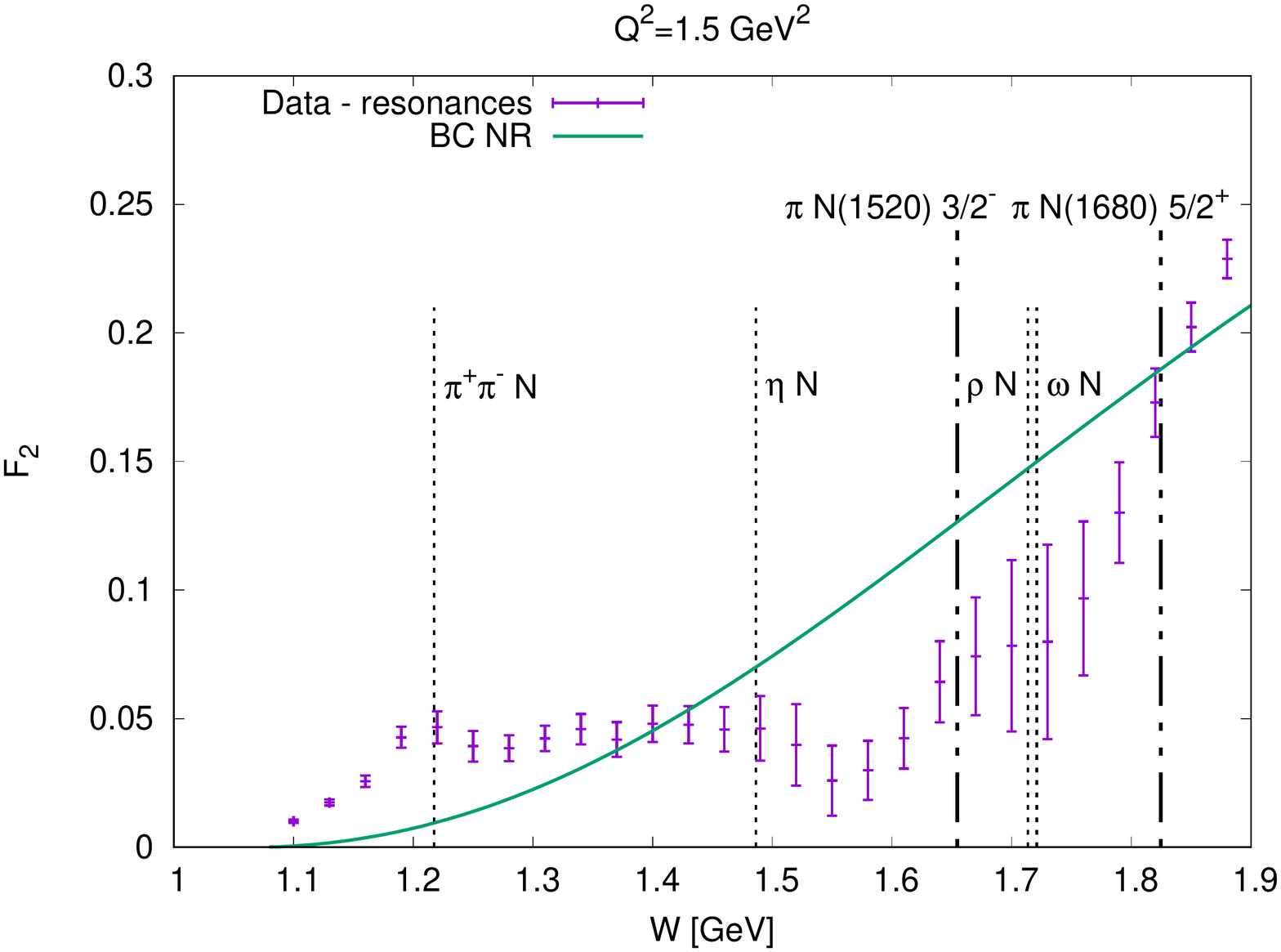}
\includegraphics[width=0.45\textwidth]{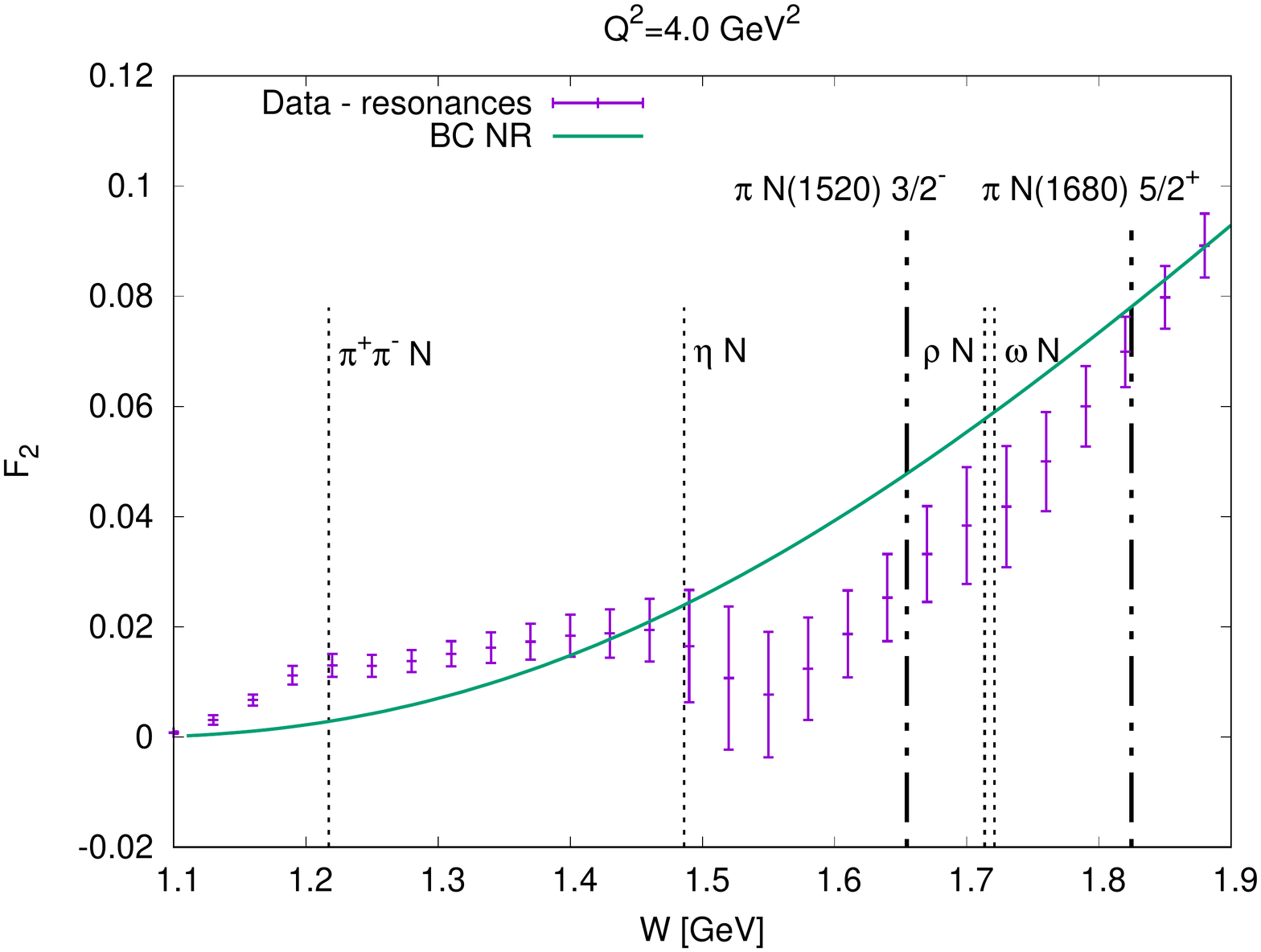}
\caption{Pseudo-data (purple points) for the differences between the $F_2$ structure function data~\cite{CLAS:SFDB,Golubenko:2019gxz} and the estimated resonant contributions, compared to the background used in Ref.~\cite{Christy:2007ve} (green curve). The dotted vertical lines show the opening of meson-nucleon electroproduction channels. The dash-dotted vertical lines show the opening of $\pi~N^*$ channels.}
\label{F:F2NR}
\end{figure*}

\begin{figure*}
\centering
\includegraphics[width=0.55\textwidth]{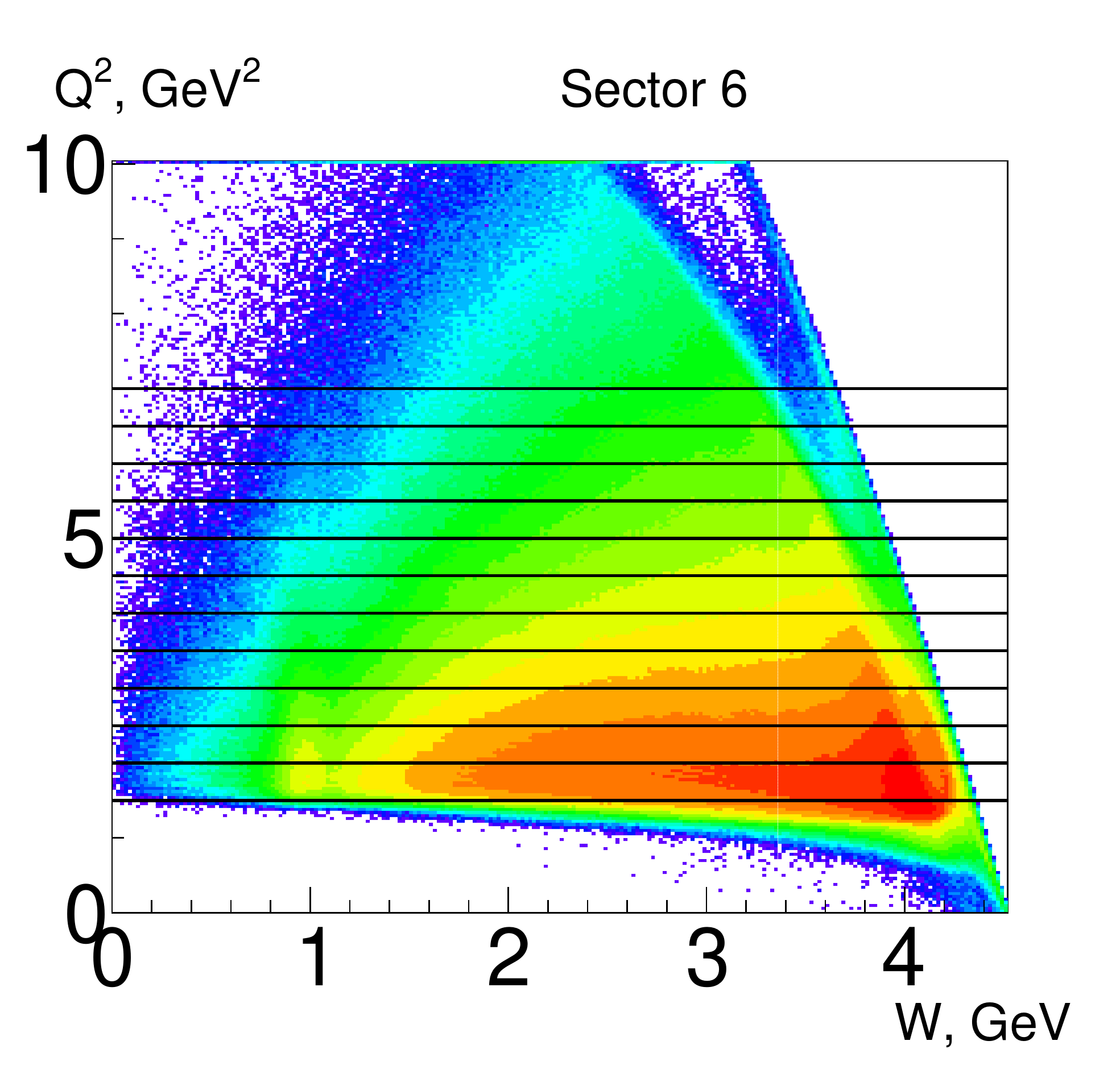}
\includegraphics[width=0.55\textwidth]{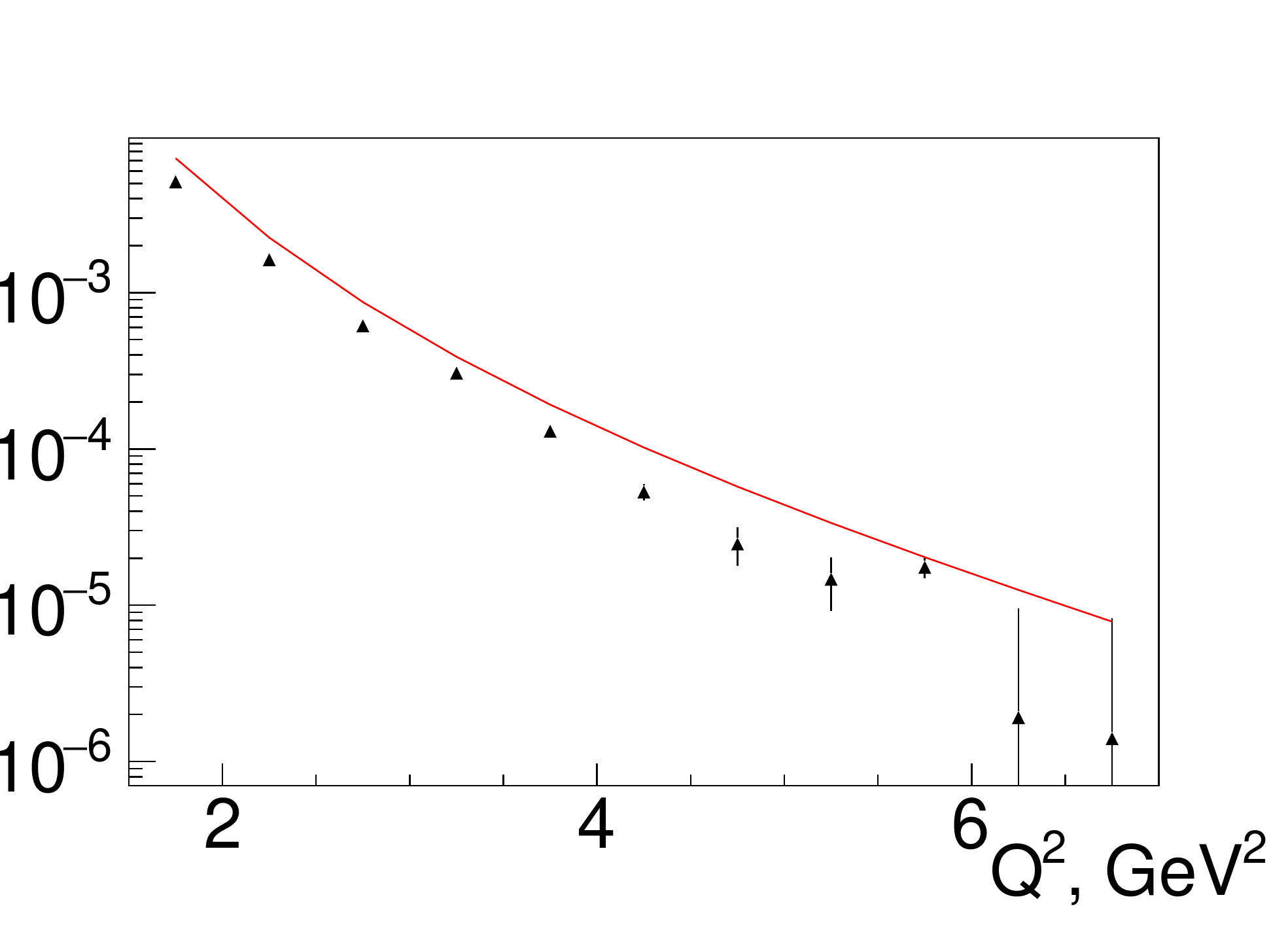}
\caption{Preliminary $Q^2$ vs $W$  distribution of inclusive electron scattering events measured with the CLAS12 detector (top). Normalized yield (arb. units) as a function of $Q^2$ (bottom) at $W$=1.22 GeV in comparison with parameterization developed in \cite{Golubenko:2019gxz}}
\label{F:clas12}
\end{figure*}

Since our evaluation of the resonant contributions depends only on exclusive electroproduction data, it enables to give a first estimate of the non-resonant contributions to the inclusive observables. This estimate consists in determining the difference between the inclusive data and the resonant contributions estimated independently from exclusive meson electroproduction data. This is shown in Fig.~\ref{F:F2NR}, and can be interpreted as pseudo-data for the background of the structure function $F_2$ measured by CLAS~\cite{Osipenko:2003bu}. Both the data and the model uncertainties are propagated into the final error bars shown. 
We also show the comparison of the contributions inferred from the data with the background model from Ref.~\cite{Christy:2007ve}. The non-resonant contributions determined by realistically accounting for the nucleon resonances demonstrate several structures and a sharp increase with $W$ from 1.6~GeV to 1.7~GeV seen in all $Q^2$-bins. One can observe several kinks in the $W$ dependence of the background for $F_2$. It appears, however, that each of them is associated with the opening of a meson-baryon channel, namely $\pi\pi N$ at 1.21~GeV, $\eta N$ at 1.49~GeV,  and $\omega N$ at 1.72~GeV. We also show the $W$ values for the opening of the $\pi^+ N(1520)~3/2^-$, $\pi^+ N(1680)~5/2^+$ and $\rho N$ channels, calculated at the resonance central masses. 
Because of the  appreciable decay widths ($\gtrsim 100$~MeV) of the unstable final states, instead of kinks these channels produce sharp growths in the $W$ dependence of the $F_2$ structure function at $1.6$~GeV~$\lesssim W \lesssim 1.7$~GeV, seen in all $Q^2$ bins. These features are not presented in the extrapolation of the non-resonant contributions from the deep inelastic scattering region into the resonance region in Ref.~\cite{Christy:2007ve}. Therefore, the realistic estimate of the resonance contributions from the experimental electrocoupling data may have impact on the insight into the parton distributions in the ground nucleon at Bjorken variable $x$ values close to unity, within the resonance region. This is the subject of our future work.

Preliminary results on inclusive electron scattering with a 10.6~GeV electron beam have recently become available from the CLAS12 detector. The achieved kinematic coverage is shown in Fig.~\ref{F:clas12} (top). The $Q^2$ dependence of the normalized inclusive event yield shown in Fig.~\ref{F:clas12} (bottom) is in good agreement with the expectation from the interpolation of the CLAS and world data on inclusive electron scattering~\cite{Golubenko:2019gxz}. The approach presented in these proceedings provides a phenomenological tool for the analyses of the CLAS12 data on inclusive electron scattering at high photon virtualities of $Q^2>5.0$~GeV$^2$ in the resonance region.

\section{Summary}\label{S:sum}
In this work, for the first time the resonant contributions to unpolarized electron scattering observables have been evaluated in the first three resonance regions, using as input the CLAS data on resonance electrocouplings~\cite{Aznauryan:2011qj,Mokeev:2018zxt}. For all three resonance peaks, we observed substantial contributions from the resonance tails in the neighboring regions. The behaviour in the $Q^2$ evolution of the resonance structures is non-trivial: the first and third  peaks decrease strongly with $Q^2$, while the second region has a softer decrease: in fact, its ratio with respect to the background stays relatively constant. These distinctively different structural features underline the importance of the separate studies of the nucleon resonances for the exploration of strong QCD.

The comparison between the resonant contributions and data from CLAS~\cite{Osipenko:2003bu} allowed to give an estimate for the separate contribution of the background at different values of $Q^2$. The electrocouplings of nucleon resonances in an even broader mass and $Q^2$ range will soon become available~\cite{Isupov:2017lnd,Fedotov:2018oan}, and with the CLAS12 endeavour on both exclusive and inclusive electron-induced reactions, the data base will be extended to the largest $Q^2$ interval ever achieved, from 0.05~GeV$^2$ to 12~GeV$^2$. The present study can thus also be used as a phenomenological tool for the analysis of CLAS12 data. Ultimately, our goal is to model resonant and background contributions in a combined fashion in wide $Q^2$ regions, in order to describe the data at low and high energies. By using the experimental results on the nucleon resonance electrocouplings we are also planning on gaining insight into the parton distributions in the ground nucleons at Bjorken variable $x$ values close to unity, within the resonance region.

\begin{acknowledgement}
This work was supported by the Deutsche Forschungsgemeinschaft (DFG, German Research Foundation), in part through the Collaborative Research Center [The Low-Energy Frontier of the Standard Model, Projektnummer 204404729 - SFB 1044], and in part through the Cluster of Excellence [Precision Physics, Fundamental Interactions, and Structure of Matter] (PRISMA$^+$ EXC 2118/1) within the German Excellence Strategy (Project ID 39083149). It was also supported by the U.S. Department of Energy, Office of Science, Office of Nuclear Physics under contract DE-AC05-06OR23177 and Skobeltsyn Nuclear Physics Institute and the Physics Department at Moscow State University.
\end{acknowledgement}

\interlinepenalty=10000
\bibliography{resonance}{}

\end{document}